\renewcommand{\Re}{\mathop{\rm Re}\nolimits}
\renewcommand{\Im}{\mathop{\rm Im}\nolimits}
\documentstyle[12pt,amssymb]{article}
\begin{document}
\begin{titlepage}
\begin{flushright}
UM-TH-96-02\\
February 1996\\
\end{flushright}
\vskip 2cm
\begin{center}
{\large\bf Model-independent Representation of Electroweak Data}
\vskip 1cm
{\large Robin G. Stuart}
\vskip 1cm
{\it Randall Physics Laboratory,\\
 University of Michigan,\\
 Ann Arbor, MI 48190-1120,\\
 USA\\}
\end{center}
\vskip .5cm
\begin{abstract}
General model-independent expressions are developed for the polarized
and unpolarized cross-sections for $e^+e^-\rightarrow f\bar f$ near
the $Z^0$ resonance. The expressions assume only the analyticity of $S$-matrix
elements. Angular dependence is included by means of a partial wave expansion.
The resulting simple forms are suitable for use in fitting data or in
Monte Carlo event generators. A distinction is made between model-independent
and model-dependent QED corrections and a simple closed expression is given for
the effect of initial-final state bremsstrahlung and virtual QED corrections.
\end{abstract}
\vskip 4cm
\end{titlepage}

\setcounter{footnote}{0}
\setcounter{page}{2}
\setcounter{section}{0}
\newpage

\section{Introduction}

The purpose of an experiment such as LEP is two-fold. First and foremost it
should measure and record experimental results with a minimum of
theoretical input or prejudice and without presupposing that the data
is described by a particular theoretical model. Only in this way can the
hard-won experimental data be of use should our present understanding of
the physics change or the Standard Model be supplanted.

The second purpose is to test the correctness of the various candidate
theoretical models that describe the physics of the processes involved.
Such theoretical models will contain parameters, such as $\sin^2\theta_W$,
that can be extracted by fitting the data with predictions of the model.
A given model is ruled out when the values obtained for the extracted
parameters differ depending on the measurement or physical
process from which it was obtained. The values of model parameters
should be extracted and recorded for comparison between past and future
experiments. However these parameters may become meaningless once the model
to which they pertain is ruled out or modified. They therefore cannot
perform the function of recording experimental data for posterity.

While the high-energy physics community has been saturated with analyses
that confront the Standard Model and its possible extensions with the
experimental data, much less attention has been given to preserving the data
in an unambiguous model-independent form.

One way to do this would be to make a complete set of raw data available.
The sheer volume of data makes this impractical. In addition, `raw' data
is seldom truly raw having been subjected to on-line triggering.
It therefore bears the stamp of the on-line selection procedure.

Another possible way is that experimentalists provide plots of raw
cross-sections as a function of, say, centre-of-mass energy or
scattering angle. Such cross-sections again suffer from the problem that
they have been subjected to on-line selection. To produce such plots,
experimental cuts generally need to be introduced.

Both of the above possibilities require that the potential user, who
wishes to test a given model for consistency with experiment, have a
fairly sophisticated machinery in place for treating the QED and other
background effects. Details of experimental cuts, detector geometry etc.\
should all be meticulously recorded.

Fortunately a third possibility exists by which the experimental results
can forever be recorded in a way that makes them straightforwardly available
for testing theoretical models as they appear. That is to identify and extract
the model-independent physical observables that are inherent in the data and
record those. The resulting set of physical parameters is small and
convenient to use having all detector-dependent effects removed from it.
In order to test the consistency of a candidate model one needs only
calculate the given physical observable in terms of the parameters
of that model and compare it to the recorded value. There is no convention-,
model- or scheme-dependence in the physical observables so that the
comparison can be safely and unambiguously made.

Part of the reason why experimental data has not generally been recorded
in this way may be the lack of understanding of the distinction between
model-independent physical observables and parameters specific to a given
model. The distinction between the two is considered in section 2.

To be fair to experimentalists, the results of LEP experiments are extracted
in a relatively model-independent way. Semi-empirical expressions are
fitted to the data in order to extract quantities such as the mass, total and
partial width of the $Z^0$ boson\cite{Blondel}. The shortcoming is that
these expressions are indeed semi-empirical and at some level of accuracy
they will fail to describe the data correctly. Being semi-empirical they
are also somewhat arbitrary and one needs to have the detailed expressions
that were used in the data analysis in order to interpret the experimental
results.

The first serious attempt to consistently describe LEP data in a
fundamentally model-independent way was made by
Borrelli {\it et al.}\cite{Borrelli}. They clearly recognized the inadequacies
inherent in model-dependent analyses of the data. Their approach involved
expressing cross-sections to ${\cal O}(\alpha)$ in terms of five independent
physical observables, $M$, $\Gamma$, $B$, $R$ and ${\cal I}$.
These observables represent the mass, width, branching ratio, non-resonant
and absorptive pieces of the matrix element respectively.
This work was extended by Isidori \cite{Isidori}. An important consequence
of ref.\cite{Borrelli} is the clear statement of the need
for five independent measurements to fully describe LEP data. Motivated
by this Consoli and Piccolo \cite{Consoli} suggested that the
final LEP scan at the $Z^0$ be extended to include five energies rather than
just three.

The shortcoming of the analysis performed in ref.\cite{Borrelli} is that
at a certain level of accuracy it becomes unclear as to just what the
physical observables are. These authors followed the conventional wisdom
and expanded about a real mass $M$ chosen to coincide with the renormalized
mass in the popular on-shell renormalization scheme. The choice of this $M$
as an expansion point is arbitrary and hence the physical observables
extracted using it will also be arbitrary.

To elucidate the difficulties of defining physical observables further,
consider the problem of determining the total
width of the $Z^0$ resonance. In principle an energy scan can be performed
for the cross-section for $e^+e^-$ producing some final state determined.
The full width at half maximum of the resulting resonance curve can then
be read off. A raw resonance curve will, of course, wear a radiative tail
generated by initial-state photon radiation. That being removed the exact
shape of the resonance curve will depend, via final-state vertex corrections,
on which final state has been selected for the measurement.
Thus the width of the resonance curve does not provide a way of directly
determining a unique model-independent total width for the $Z^0$ boson.

The mass is equally problematic to define. Even in the most na\"\i ve
of analyses the resonance peak lies far from what is assumed to be the
mass. A discussion of the issues involved can be found in ref.\cite{Stuart3}
and in the following section.

As discussed in ref.\cite{Stuart3} similar problems exist for the definition
of partial widths.

Fortunately a way does exist to define
the physical properties of the $Z^0$ boson and describe LEP data
in a simple and truly model-independent way. Prior to 1991 most calculations
of physics at the $Z^0$ resonance where demonstrably gauge-dependent. It was
shown in ref.\cite{Stuart1} how the gauge-dependence could be removed by
appealing to the known properties of the analytic $S$-matrix near resonance.
The solution involved starting from the known structure of the complete
$S$-matrix element and then performing a Laurent expansion about its complex
pole, $s_p$. It was pointed out there, and independently in
ref.\cite{Willenbrock}
that the physical mass, traditionally used for unstable particles
and defined from $S$-matrix theory, differed significantly
from that being extracted by LEP. The analytic $S$-matrix seems to provide
the only way of defining the properties of the $Z^0$ boson in a simple
and truly model-independent way. As such it is the most appropriate and
robust way of preserving LEP data.

In ref.\cite{Stuart2}, a paper concerned with the general renormalization
of the pole expansion, it was shown that the pole expansion could be used to
obtain a simple general expression for the $S$-matrix element
near the $Z^0$ resonance for
$e^+e^-\rightarrow f\bar f$, with $f$ being a generic fermion species.
Provided one is not too close to a production threshold the general
matrix element takes the form of a Laurent expansion
\begin{equation}
A(s)=\frac{R}{s-s_p}+\sum_{n=0}^\infty B_n (s-s_p)^n
\label{eq:FullLaurentExp}
\end{equation}
for fixed scattering angle. It was clearly stated that this expression
was applicable to two-particle final states thereby excluding
bremsstrahlung diagrams. It easily follows that neglecting terms
${\cal O}(\Gamma^2_Z/M^2_Z)$ and higher
\begin{equation}
A(s)=\frac{R}{s-s_p}+B_0
\label{eq:OalphaLaurentExp}
\end{equation}
and therefore depends on three complex numbers, $s_p$, $R$ and $B_0$.
Here and in what follows ${\cal O}(\Gamma_Z/M_Z)\equiv{\cal O}(N_f\alpha)$
where $N_f$ is the number of fermions species into which the $Z^0$ can decay.
The cross-section will thus depend on five real parameters, in agreement
with Borrelli {\it et al.}\cite{Borrelli}, since the overall phase is lost.
The difference here is that expansion is made about the pole which is
a fundamental property of the $S$-matrix element. The resulting coefficients
and definitions of physical observables are therefore not dependent on an
arbitrary choice of real expansion point $M$.
If it were not for the tree-level photon exchange diagram
we could drop $B_0$ in eq.(\ref{eq:OalphaLaurentExp}) to obtain the stated
level of accuracy. $A(s)$ would then depend
on 3 real numbers only. The question of the model-independent
parameterization of the angular dependence of the scattering amplitude
was not considered in ref.\cite{Stuart2}.

Subsequently Leike {\it et al.} \cite{Leike} repeated the analysis of
ref.\cite{Stuart2} explaining how the Laurent expansion could be
implemented in practice. They carried out an actual fit to data including
QED corrections. They suggested a parameterization that made the
presence of photon exchange diagrams explicit rather than absorbing them
into background,
\begin{equation}
A(s)=\frac{R_Z}{s-s_p}+\frac{R_\gamma}{s}+B(s)
\label{eq:RiemannExp}
\end{equation}
where $B(s)$ is a function having no poles.
This parameterization was incorporated into the computer program
{\tt SMATASY} \cite{KirschRiemann}.

Including the photon contribution $R_\gamma/s$ and background together
in this way may lead to difficulties however. Whereas
eq.(\ref{eq:FullLaurentExp}) is a self-consistent Laurent expansion about
a simple pole, $s_p$, and valid within some radius of convergence,
eq.(\ref{eq:RiemannExp}) is not. Hence the coefficients $R_\gamma$ and
$B_i$ are not independent quantities. In particular the photon exchange
term may be written as a Taylor series expansion about $s_p$,
\begin{equation}
\frac{R_\gamma}{s}=\frac{R_\gamma}{s_p}
                   -\frac{R_\gamma}{s_p^2}(s-s_p)
                   +\frac{R_\gamma}{s_p^3}(s-s_p)^2+...
\end{equation}
Thus any change in $R_\gamma$ can be exactly compensated by a corresponding
change in the coefficients $B_i$ which is an undesirable feature for
fitting. Whereas finite
truncations of the series $\sum B_i(s-s_p)^i$ may produce adequate fits,
as more and more terms are included in the series the coefficients become
indeterminate.

The L3 collaboration \cite{L3} performed an analysis of their data based on
eq.(\ref{eq:RiemannExp}), curiously without a citation to ref.\cite{Stuart2}
that was the original source for the $S$-matrix approach.
They truncated their amplitude at
\begin{equation}
A(s)=\frac{R_Z}{s-s_p}+\frac{R_\gamma}{s}
\label{eq:L3Exp}
\end{equation}
so that the coefficients are indeed independent and thus amenable to
fitting.

In this paper we look in more detail at the model-independent $S$-matrix
description of LEP data as a way of preserving the experimental results
in a transparent and natural way that will continue to be understandable
and useful for many years to come. In section 3 the basic $S$-matrix
formalism is reviewed and extended. It is shown how to describe the angular
dependence of the scattering amplitude in a model-independent way.
In section 4 the inclusion of QED corrections is discussed. A distinction
is made between model-independent and model-dependent bremsstrahlung.
We also give a simple exact formula for initial-final state interference
corrections to the resonant term.

\section{Physical observables vs. Model-dependent parameters}

In this paper the term model-independent physical observable will be taken
to mean a quantity that can be directly defined in terms of some set of
experimental measurements without the need for input from some theoretical
model. Thus the electromagnetic coupling constant, $\alpha$, is exactly
defined from the result of a Thomson scattering experiment,
\begin{equation}
\sigma_T=\frac{8\pi}{3}\frac{\alpha^2}{m_e^2}.
\end{equation}

Similarly the muon decay constant, $G_\mu$, is exactly defined from the
the experimental measurement of the muon lifetime, $\tau_\mu$ through the
relation,
\begin{equation}
\tau_\mu^{-1}=\frac{G_\mu^2m_\mu^5}{192\pi^3}
                \left(1-\frac{8m_e^2}{m_\mu^2}\right)
                \left[1+\frac{3}{5}\frac{m_\mu^2}{M_W^2}
                           +\frac{\alpha}{2\pi}\left(\frac{25}{4}-\pi^2\right)
                            \left(1+\frac{2\alpha}{3\pi}\ln\frac{m_\mu}{m_e}
                            \right)
                    \right]\label{eq:muonlifetime}
\end{equation}
The relation (\ref{eq:muonlifetime})
is quite complicated but nevertheless, once $\tau_\mu$ is
measured, $G_\mu$ is unambiguously defined. The complexity of
(\ref{eq:muonlifetime}) arises from
an attempt to factor out QED corrections. There may exist a more convenient
or pragmatic way of defining $G_\mu$ from $\tau_\mu$ but the one given is
well-established and in common use. Note that, in principle at a certain
level of precision, what one means by the lifetime of an unstable particle
becomes unclear because the decay curve is not precisely exponential.
These considerations are relevant for the $Z^0$ boson but,
because of its extremely long lifetime, are unlikely to ever be of concern
for the muon.

The position of the pole, $s_p$, may be regarded as a model-independent
physical observable because its existence depends only on the analyticity
of the $S$-matrix which is ultimately believed to derive from
causality. Its value can, in principle, be extracted from measurements
of the cross section $\sigma(e^+e^-\rightarrow f\bar f)$ over a large energy
range using analytic continuation onto the second Riemann sheet.
The same value of $s_p$ will be obtained for any process involving an
intermediate $Z^0$.

In a similar way the residue $R_{if}$ at the pole for a given $f\bar f$
final state can be extracted from experiment without detailed
model-dependent input. It is known to factorize, $R_{if}=R_i\cdot R_f$ and
the $R_f$ can form the basis for a model-independent definition of the
partial width \cite{Stuart3}.

The essential point about a model-independent physical observable is that
once a set of experimental measurements is available its value is fixed.
In the case of Thomson scattering or the measurement of the muon lifetime,
a single number is returned by the experiment and what
one means by a model-independent physical observable is clear-cut.
Things become less obvious for observables, such as $s_p$ and $R_{if}$,
that need to be extracted by fitting experimental data over a certain
energy range but they still represent viable model-independent
physical observables.

By contrast, model-dependent parameters require the lagrangian of the
underlying model be known and specified. A detailed calculation is required
in order to fit the experimental data. The values obtained will be sensitive
to which renormalization scheme was used in the calculation and will be
subject to what {\it ad hoc\/} modifications (improved Born approximations,
effective mixing angles and the like) one makes above and beyond a
consistent truncated perturbation series.

A good example of a model-dependent parameter is $\sin^2\theta_W$. From a
theoretical point of view, in the Standard Model, $\theta_W$ is the
angle of rotation that diagonalizes the mass matrix of the neutral $W_3$
and $B$ boson that appears in the lowest-order renormalized lagrangian.
In all renormalization schemes the relation
\begin{equation}
\sin^2\theta_W=1-{M_W^2\over M_Z^2}\label{eq:Sirlinsin}
\end{equation}
holds provided $M_W$ and $M_Z$ are the renormalized masses in the
particular scheme that has been chosen. In the $\overline{\rm MS}$
renormalization scheme $M_W$ and $M_Z$ depend on an arbitrary scale and
hence so does $\sin^2\theta_W$. Once a particular renormalization scheme
has been chosen, experimental results may be used as input to determine
the values of the renormalized parameters. These renormalized parameters
generally have no physical meaning outside of the particular model or
renormalization scheme that has been chosen and are eminently unsuitable
for recording experimental results. They can, however, still be used to
test a given model by using it to make predictions for other physical
observables. Thus the $\overline{\rm MS}$ renormalization scheme has
renormalized masses that are clearly unphysical but is still a viable and
convenient scheme to use in many situations.

At the risk of blurring the distinction between physical quantities and
(unphysical) renormalized parameters, one can try to define a
renormalization scheme that sets the renormalized parameters to be
equal to physical observables as was done in ref.s\cite{Sirlin}.
Again the problem of just what are the physical observables arises.
Furthermore in gauge theories, with their interrelated
coupling-constant and mass counterterms, one must take care not to
violate Ward identities. In its original incarnation the on-shell
renormalization scheme \cite{Sirlin} used a definition for the
physical mass that was subsequently shown \cite{SirlinMass1,SirlinMass2}
to be gauge-dependent. In the same works it was suggested to modify
the on-shell renormalization scheme in such a way that the renormalized
mass is identified with the manifestly gauge-invariant,
but arbitrary, quantity (\ref{eq:SirlinMass})
constructed from the pole of the $S$-matrix.
In principle, since the renormalized mass
is not a physical quantity, it is not required to be gauge-invariant
{\it a priori\/} although it is of great practical convenience to have
it so. Consistency of this renormalization scheme with Ward identities
has yet to be explored.

It is important to keep in mind
the clear distinction between the physical mass and the unphysical
renormalized mass. The former is a model-independent physical observable
but the latter is not.
An unstable particle is associated with a pole in the $S$-matrix element
$s=s_p$ lying below the real axis. It is this complex number as a whole
that is physically significant. For convenience two real numbers can be
extracted from $s_p$ and identified with the mass, $M$, and width, $\Gamma$.
There is no fundamental way of doing this and any such decomposition will be
arbitrary. Two long-standing conventions are to define
\begin{eqnarray}
s_p&=&M^2-iM\Gamma,\label{eq:firstmass}\\
s_p&=&\left(M-\frac{i}{2}\Gamma\right)^2\label{eq:secondmass}.
\end{eqnarray}
It was noted \cite{Stuart1,Willenbrock} that in the case of the $Z^0$
the definition (\ref{eq:firstmass}) produces a value that is 34\,MeV below
the value being extracted by LEP. The latter is based on the use of the
on-shell renormalization scheme. In ref.\cite{SirlinMass1}
it was suggested that, rather than employ the traditional definitions,
(\ref{eq:firstmass}) and (\ref{eq:secondmass}),
the $Z^0$ boson mass should be defined as
\begin{equation}
M_Z^2=\Re s_p+\frac{(\Im s_p)^2}{\Re s_p}\label{eq:SirlinMass}
\end{equation}
This is reasonable because it turns out to be numerically close to the value
being extracted by LEP and hence requires minimal modification of existing
analyses and existing experimental results. However it is no more nor less
fundamental than (\ref{eq:firstmass}) or (\ref{eq:secondmass}) and no
more nor less deserving of the title of physical mass.

Attempts have been made to give physical definitions to $\sin^2\theta_W$.
Llewellyn-Smith and Wheater \cite{LlewellynWheater} defined an experimental
$\sin^2\theta_W^{\rm exp}$ from the ratios of charged- and neutral-current
neutrino scattering experiments.
Although this definition constitutes a model-independent physical observable
it is not the parameter that appears in the Standard Model lagrangian.
It is really some convenient way of encapsulating the result of cross-section
measurements that once extracted from experiment must be
corrected by means of a detailed model-dependent calculation,
in order to yield a value for $\sin^2\theta_W$ in some particular
renormalization scheme.

\section{Model-independent lineshape}

The general matrix element for the process $e^+e^-\rightarrow f\bar f$
near resonance is a sum over current-current interactions
\begin{equation}
A(s,t)=\sum_{i,f} {\cal M}_{if}(s,t) J_i\cdot J_f
\end{equation}
and is a function of the usual Mandelstam variables $s$ and $t$.
The form factors ${\cal M}_{if}(s,t)$ are analytic functions of $s$ and $t$.
For massless electrons and final-state fermions only vector and axial-vector
currents can appear and thus
\begin{equation}
A(s,t)=\sum_{i,f=L,R} {\cal M}_{if}(s,t)
       [\bar v(p_{e^+})\gamma_\mu\gamma_i u(p_{e^-})]
       [\bar u(p_{f})\gamma_\mu\gamma_f v(p_{\bar f})]
\label{eq:Aif}
\end{equation}
where $u$ and $v$ are the fermion wave functions and $\gamma_L$, $\gamma_R$
are the usual helicity projection operators $\gamma_{L,R}=(1\pm\gamma_5)/2$.
For massive final-state fermions, magnetic moment terms,
$\sigma_{\mu\nu}q^\nu\gamma_L$ and $\sigma_{\mu\nu}q^\nu\gamma_R$,
are possible but these can be at most
${\cal O}(\alpha m_f^2)$ in the final cross-section and will be dropped in
the later analysis. We will therefore discard them at the outset.
At this point it is convenient to go to the centre-of-mass frame and express
${\cal M}_{ij}$ as a function of $s$ and $\cos\theta$ the cosine of the
scattering angle. These may be expanded as a Laurent series about the
complex pole of the scattering amplitude
\begin{equation}
{\cal M}_{if}(s,\cos\theta)=\frac{R_{if}}{s-s_p}+B_{0,if}(\cos\theta)
                                +B_{1,if}(\cos\theta)(s-s_p)+...
\label{eq:Mif}
\end{equation}
that is valid within a radius of convergence defined by the position of the
nearest branch point which corresponds to a production threshold.
For the $Z^0$ resonance the low-order thresholds for fermion production lie
sufficiently far from the resonance so as to be unlikely to ever to be of
concern. However some interesting physical consequences arise in the case of
nearby thresholds \cite{Stuart3,BhattaWillen}.

As a consequence of Fredholm theory it is known that
the residue at the pole factorizes and we may write $R_{if}=R_i\cdot R_f$
where $R_i$ does not depend on the properties of the final-state particle
and $R_f$ is independent of those of the initial-state particle.
The functions $B_{n,if}$ can be expanded in partial waves,
\begin{equation}
B_{n,if}(\cos\theta)=\sum_{m=0}^\infty B^{nm}_{if}P_m(\cos\theta)
\end{equation}
where $P_m(x)$ is the Legendre polynomial of order $m$.
All constants $s_p$, $R_{if}$ and $B^{nm}_{if}$ are, in principle,
complex numbers.

The general differential cross-section in the centre-of-mass frame for the
process $e^+e^-\rightarrow f\bar f$ with massless incoming electrons of
polarization, $P$, colliding with unpolarized positrons is
\begin{equation}
\frac{d\sigma}{d\Omega}=\left(\frac{1+P}{2}\right)^2
                        \left(\frac{d\sigma_{LL}}{d\Omega}
                             +\frac{d\sigma_{LR}}{d\Omega}\right)
                        +\left(\frac{1-P}{2}\right)^2
                        \left(\frac{d\sigma_{RR}}{d\Omega}
                             +\frac{d\sigma_{RL}}{d\Omega}\right)
\end{equation}
with
\begin{equation}
\frac{d\sigma_{ij}}{d\Omega}=\frac{s\beta}{64\pi^2}\left\{
                        \left(\frac{1\pm\beta\cos\theta}{2}\right)^2
                        \left\vert{\cal M}_{if}\right\vert^2
                       +\left(\frac{1-\beta^2}{4}\right)
                        \Re{\cal M}_{if}{\cal M}_{i,-f}^*\right\}
\end{equation}
where $\beta=\sqrt{1-4m_f^2/s}$ and $m_f$ is the mass of the final state
fermion.
The upper sign pertains for $i=f$ and the lower for $i\ne f$. The second
term is a helicity flip term and is needed only for massive final-state
fermions. There the index $-f$ means the opposite helicity to $f$.
Performing the angular integrations over the full solid angle it becomes
clear that cross-sections must take the general form
\begin{equation}
\sigma_{if}=\frac{s\beta}{32\pi}\left\{
       \frac{c^{if}_{-2}}{\vert s-s_p\vert^2}
       +\Re\left(\frac{c^{if}_{-1}}{s-s_p}\right)+c^{if}_0\right\}
\label{eq:generalxsec}
\end{equation}
keeping terms up to order ${\cal O}(\Gamma^2_Z/M_Z^2)$ relative to the leading
one and the leading term proportional to
$1-\beta^2=4m_f^2/s$. This last term only contributes in the case of the
$b$-quark. The above form is likely to be adequate for all practical purposes
in the foreseeable future.
The constants $c^{if}_{-2}$ and $c^{if}_0$ are real and $s_p$ and
$c^{if}_{-1}$ are complex and hence to this level of accuracy the
cross-section depends only on six real constants,
$\Re s_p$, $\Im s_p$, $c^{if}_{-2}$, $\Re c^{if}_{-1}$, $\Im c^{if}_{-1}$
and $c^{if}_0$. Because $s_p$ is the same for all $\sigma_{if}$, summation
over initial- or final-states will lead to a cross-section of the same
overall form.
Using the usual normalization for the Legendre polynomials
\[
\int_{-1}^{1}[P_n(x)]^2\,dx=\frac{2}{2n+1}
\]
and the results

\[
\int_{-1}^{1}\left(\frac{1\pm\beta x}{2}\right)^2 P_n(x)\,dx
        =\left\{\begin{array}{rr}
              \frac{\displaystyle3+\beta^2}{\displaystyle6},& n=0\\
                   & \\
              \pm\frac{\displaystyle\beta}{\displaystyle3},&    1\\
                   & \\
              \frac{\displaystyle\beta^2}{\displaystyle15},&    2
              \end{array}
         \right.
\]
we have
\begin{eqnarray}
c_{-2}^{if}&=&\frac{3+\beta^2}{6}\vert R_{if}\vert^2
       +\left(\frac{1-\beta^2}{4}\right)\Re R_{if}R_{i,-f}^*\label{eq:cifm2}\\
c_{-1}^{if}&=&2R_{if}\left(\frac{3+\beta^2}{6}{B_{if}^{00}}^*
                    \pm\frac{\beta}{3}{B_{if}^{01}}^*
                    +\frac{\beta^2}{15}{B_{if}^{02}}^*\right)\nonumber\\
      &+&2R_{if}(s_p-s_p^*)\left(\frac{3+\beta^2}{6}{B_{if}^{10}}^*
                    \pm\frac{\beta}{3}{B_{if}^{11}}^*
                    +\frac{\beta^2}{15}{B_{if}^{12}}^*\right)\label{eq:cifm1}\\
c_0^{if}&=&2R_{if}\left(\frac{3+\beta^2}{6}{B_{if}^{10}}^*
                 \pm\frac{\beta}{3}{B_{if}^{11}}^*
                 +\frac{\beta^2}{15}{B_{if}^{12}}^*\right)\nonumber\\
      &+&\int_{-1}^1\,d(\cos\theta)
                 \left(\frac{1\pm\beta\cos\theta}{2}\right)^2
                           \vert B_{0,if}(\cos\theta)\vert^2
\label{eq:cif0}
\end{eqnarray}
Using the properties of Legendre polynomials
\begin{eqnarray}
\int_{-1}^1&&{}\!\!\!\!\!\!\!\!\!\!\!\!d(\cos\theta)
                 \left(\frac{1\pm\beta\cos\theta}{2}\right)^2
                           \vert B_{0,if}(\cos\theta)\vert^2\nonumber\\
&=&\frac{1}{2}\sum_{m=0}^\infty\left(\frac{1}{(2m+1)}
                           +\beta^2\frac{(2m^2+2m-1)}{(2m-1)(2m+1)(2m+3)}
                             \right)\vert B_{if}^{0m}\vert^2\nonumber\\
 &\pm&\beta\sum_{m=0}^\infty\frac{(m+1)}{(2m+1)(2m+3)}
                            \Re B_{if}^{0m}{B_{if}^{0,m+1}}^*\nonumber\\
 &+&\beta^2\sum_{m=0}^\infty\frac{(m+1)(m+2)}{(2m+1)(2m+3)(2m+5)}
                            \Re B_{if}^{0m}{B_{if}^{0,m+2}}^*
\end{eqnarray}

To order ${\cal O}(\Gamma_Z/M_Z)$ the last term in eq.(\ref{eq:generalxsec})
can be dropped and the cross-section then depends on five real constants
as first pointed out by Borrelli~{\it et al.}\ \cite{Borrelli}. This same total
was obtained in ref.\cite{Stuart2} using arguments based on analyticity
similar to those above.
The form (\ref{eq:generalxsec}) is extremely general
and robust being valid for polarized and unpolarized matrix elements as well
as for angular integrations over less than the full solid angle.

To this level of accuracy one further assumption can be made to
further simplify the expressions (\ref{eq:cifm2})--(\ref{eq:cif0}). Since
no new particles have been found with a mass $m\lesssim M_Z/2$ we may be sure
that the first terms that give rise to corrections that depend on the
scattering angle $\theta$ are box Feynman diagrams corresponding to
two $W$ or two $Z$ exchange. These diagrams are one-loop and non-resonant.
They therefore represent corrections of ${\cal O}(\alpha\Gamma_Z/M_Z)$
relative to the lowest order and may be dropped by setting $B_{if}^{1n}=0$.

The constants appearing in eq.~(\ref{eq:generalxsec}) can be extracted from
data in a model-independent manner and can be calculated in any theoretical
model. By comparing their measured and predicted value candidate models may
be tested. The form (\ref{eq:generalxsec}) assumes only the analyticity
of the $S$-matrix element which is a consequence of
causality. It may happen that some of the constants in the above
parameterization may seem to be poorly determined especially those in higher
orders. This is a fact of life and represents a limit on how well the
model-independent parameters can be determined from a given experiment. The
parameters in this formulation, once determined, remain valid even in the face
of dramatic changes to theoretical models. They can only be adjusted by
improved experiments and remain a fundamental and meaningful description
of the data. They are thus largely impervious even to possible profound
changes to theoretical understanding.
The temptation should be resisted to inject more detailed model-dependent
assumptions into the data analysis even when this seems to yield tighter bounds
on the parameters. If it is done it should be in addition rather, than instead
of the extraction of the above constants.

The complete generality of eq.(\ref{eq:generalxsec}) means that Monte Carlo
event generators can be set up assuming that the underlying
cross-section is of this form. The constants $s_p$, $R_{if}$ and
$B_{if}^{mn}$ corresponding to the predictions of an particular model
may then be input from independent sources and consistency with
experimental results studied. This provides an efficient method of
parameterizing corrections that depend on the scattering angle, $\theta$.
Typically these come from box diagrams whose analytic structure is
complicated and cumbersome for Monte Carlo simulations.

\section{QED Corrections}

Ultimately in the confrontation of theoretical predictions with
experimental data QED corrections must be taken into account. These QED
corrections may be grouped into two classes; model-independent and
model-dependent. Model-independent QED corrections are those in which
the photon is attached only to external fermion legs. Such corrections
sense nothing of the detailed structure of the underlying model. They
may therefore be accounted for using a structure function approach
\cite{FadinKuraev,NicrosiniTrentadue1,NicrosiniTrentadue2} applied to
a general cross-section of the form (\ref{eq:generalxsec}) or by
Monte Carlo methods. In general the Feynman diagrams
contributing to the model-independent QED corrections are infrared
divergent and can therefore give rise to large logarithms particularly if
strong cuts are applied to the photon energy.

Model-dependent QED corrections are those in which photons are connected
to internal charged particles. For example a photon that is produced by
bremsstrahlung off an internal $W$ is sensitive to the charged current
structure of the underlying model. Such corrections cannot be treated in
a model-independent manner. They are however always infrared finite and so
do not give rise to anomalously large corrections. In the case where
final-state photons are individually detected an $S$-matrix motivated
form analogous to (\ref{eq:Aif}) and (\ref{eq:Mif}) could be developed
for the process $e^+e^-\rightarrow f\bar f\gamma$. The size of the
model-dependent QED corrections represent the level of accuracy to
which a given model-independent analysis is valid.

Applying the structure function approach to treat the initial-state
QED corrections leads to a corrected cross-section in the form of a
convolution integral
\begin{equation}
\sigma_T(s)=\int ds^\prime\sigma(s^\prime)\rho_{\rm ini}(1-s^\prime/s).
\end{equation}
where $\rho_{\rm ini}$ is a known structure function.

Final-state corrections may be treated similarly but for most purposes result
in an overall multiplicative factor being applied to the cross-section.
Detailed descriptions can be found in the literature \cite{Leike,Bardin}.

QED corrections to asymmetries may also be handled in this manner
\cite{RiemannAsym}.

The initial-state and final-state QED corrections are remarkable in their
compactness and their relative simplicity. In the case of initial-final state
corrections, i.e.\ diagrams having a photon connected to both the initial-state
and final-state fermion lines, similar convolution integral forms have
been derived \cite{Bardin} however a simple exact expression exists
for first order QED corrections to the resonant part of the cross-section.
This expression can be obtained directly from refs.\cite{KuhnStuart,JKSW}.
The ${\cal O}(\alpha)$ corrections to the cross-section $\sigma_{if}(s)$
coming from
initial-final state bremsstrahlung and virtual corrections applied to the
resonant parts of the amplitude are exactly given by
\begin{equation}
\Delta\sigma_{if}(s)=\pm\frac{3\alpha}{\pi}Q_iQ_f
              I_2\left(\frac{s_p}{s},k_{\rm max}\right)\sigma^R_{if}(s).
\label{eq:xseccorrection}
\end{equation}
and
\begin{equation}
\sigma^R_{if}(s)=\frac{s\beta}{32\pi}\frac{c^{if}_{-2}}{\vert s-s_p\vert^2}.
\end{equation}
The upper sign in eq.(\ref{eq:xseccorrection})
pertains if the polarization $i=f$ and the lower if $i\ne f$.
$Q_i$ and $Q_f$ are the electric charges of the initial- and final-state
fermions. $k_{\rm max}$ is the maximum allowed photon energy in the
bremsstrahlung contribution expressed as a fraction of the centre of mass
energy, $2E_\gamma/\sqrt{s}<k_{\rm max}$.
\begin{equation}
I_2(z,k_{\rm max})=\Re\left\{z(z+1)\ln\frac{k_{\rm max}+z-1}{z}
                            +(z-1)(1-k_{\rm max})\right\}
                  -\ln|z|-2\ln k_{\rm max}
\end{equation}
For rather loose cuts on the photon energy, i.e. $k_{\rm max}$ near $1$, the
function $I_2$ passes through zero somewhere near the resonance.
This behaviour was explained
recently \cite{Stuart4} on physical grounds. Near resonance a physical
unstable $Z^0$ with a finite lifetime is created. Its finite propagation length
means that the virtual photon must itself propagate a finite distance
in order to connect the initial and final states and the amplitude is therefore
reduced. Alternatively the finite propagation length may be regarded as
resulting in a loss of correlation between initial and final states.
The upshot is that initial-final state QED corrections have a rather small
but manageable contribution to the resonant lineshape when cuts are loose.

\section{Acknowledgments}

The author wishes to thank T.\ Riemann for useful suggestions.
This work was supported in part by the U.S.\ Department of Energy.

\end{document}